\documentstyle[12pt,epsfig,a4]{article} 
\newcommand{\vs}{\vspace{-0.25cm}}
\begin{document} 

\begin{center}
{\Large{\bf Radiative corrections to neutral pion-pair production}}  
\bigskip

N. Kaiser\\
\medskip
{\small Physik-Department T39, Technische Universit\"{a}t M\"{u}nchen,
    D-85747 Garching, Germany}
\end{center}
\medskip
\begin{abstract}
We calculate the one-photon loop radiative corrections to the neutral pion-pair
photoproduction process $\pi^-\gamma \to \pi^-\pi^0\pi^0$. At leading order this
reaction is governed by the chiral pion-pion interaction. Since the chiral 
$\pi^+\pi^-\to\pi^0\pi^0$ contact-vertex depends only on the final-state 
invariant-mass it factors out of all photon-loop diagrams. We give analytical 
expressions for the multiplicative correction factor $R\sim \alpha/2\pi$ 
arising from eight classes of contributing one-photon loop diagrams. An 
electromagnetic counterterm has to be included in order to cancel the 
ultraviolet divergences generated by the photon-loops. Infrared finiteness of 
the virtual radiative corrections is achieved (in the standard way) by 
including soft photon radiation below an energy cut-off $\lambda$. The 
radiative corrections to the total cross section vary between $+2\%$ and 
$-2\%$ for center-of-mass energies from threshold up to $7m_\pi$. The finite 
part of the electromagnetic counterterm gives an additional constant 
contribution of about $1\%$, however with a large uncertainty.  
\end{abstract}
\bigskip


\section{Introduction and summary}
The pions ($\pi^+,\pi^0,\pi^-$) are the Goldstone bosons of spontaneous chiral 
symmetry breaking in QCD. Their strong interaction dynamics at low energies 
can therefore be calculated systematically (and accurately) with chiral 
perturbation theory in form of a loop expansion based on an effective chiral
Lagrangian. The very accurate two-loop predictions \cite{cola} for the S-wave 
$\pi\pi$-scattering lengths, $a_0=(0.220\pm 0.005)m_\pi^{-1}$ and $a_2=(-0.044
\pm 0.001)m_\pi^{-1}$, have been confirmed experimentally by analyzing the 
$\pi\pi$ final-state interaction effects occurring in various (rare) charged 
kaon decay modes \cite{bnl,batley,cusp}. Electromagnetic processes with pions 
offer further possibilities to test chiral perturbation theory. For example, 
pion Compton scattering $\pi^- \gamma \to\pi^- \gamma$ allows one to extract 
the electric and magnetic polarizabilities ($\alpha_\pi$ and $\beta_\pi$) of the 
charged pion. Chiral perturbation theory at two-loop order gives for the 
dominant pion polarizability difference the firm prediction $\alpha_\pi-\beta_\pi
=(5.7\pm1.0)\cdot 10^{-4}\,$fm$^3$ \cite{gasser}. It is however in conflict with 
the existing experimental results from Serpukhov $\alpha_\pi-\beta_\pi=(15.6\pm 
7.8)\cdot 10^{-4}\,$fm$^3$ \cite{serpukov} and MAMI $\alpha_\pi-\beta_\pi=(11.6
\pm 3.4)\cdot 10^{-4}\,$fm$^3$ \cite{mainz} which amount to values more than 
twice as large. Certainly, these existing experimental determinations of 
$\alpha_\pi-\beta_\pi$ raise doubts about their correctness since they violate
the chiral low-energy theorem notably by a factor 2. It is worth to note that a 
recent dispersive analysis \cite{mouss} of the Belle data for $\gamma \gamma 
\to \pi^+\pi^-$ gives the fit value $\alpha_\pi-\beta_\pi=4.7\cdot 10^{-4}\,
$fm$^3$, compatible with chiral perturbation theory.  

In that contradictory situation it is promising that the ongoing COMPASS 
experiment \cite{compass} at CERN aims at remeasuring the pion 
polarizabilities, $\alpha_\pi$ and $\beta_\pi$, with high statistics using the 
Primakoff effect. The scattering of high-energy negative pions in the 
Coulomb-field of a heavy nucleus (of charge $Z$) gives access to cross 
sections for $\pi^-\gamma$ reactions through the equivalent photon method 
\cite{pomer}. The consistent theoretical framework to extract the pion 
polarizabilities from the measured cross sections for (low-energy) pion 
Compton scattering $\pi^- \gamma \to\pi^- \gamma$ or the primary pion-nucleus 
bremsstrahlung process $\pi^- Z \to\pi^- Z \gamma$ has been described (in
one-loop approximation) in refs.\cite{picross,comptcor}. It has been stressed
that at the same order as the polarizability difference $\alpha_\pi-\beta_\pi$ 
there exists a further (partly compensating) pion-structure effect in form of 
a unique pion-loop correction (interpretable as photon scattering off the 
''pion-cloud around the pion''). In addition to these strong interaction 
effects, the QED radiative corrections to real and virtual pion Compton 
scattering $\pi^-\gamma^{(*)} \to \pi^- \gamma$ have been calculated in 
refs.\cite{comptcor,bremscor}. The relative smallness of the 
pion-structure effects in low-energy pion Compton scattering \cite{picross} 
makes it necessary to include such higher order electromagnetic corrections. The
COMPASS experiment is set up to detect simultaneously various (multi-particle) 
hadronic final-states which are produced in the Primakoff scattering of 
high-energy pions. The neutral pion production channel $\pi^-\gamma\to \pi^-
\pi^0$ serves as a test of the QCD chiral anomaly by measuring the $\gamma 
3\pi$ coupling constant $F_{\gamma  3\pi}= e/(4\pi^2 f_\pi^3) = 9.72\,$GeV$^{-3}$. For 
the two-body process $\pi^-\gamma\to \pi^-\pi^0$  the one-loop \cite{picross,
bijnens} and two-loop corrections \cite{hannah} of  chiral perturbation theory 
as well as QED radiative corrections \cite{ametller} have already been worked 
out. 

The $\pi^- \gamma$ reaction with three charged pions in the final-state is used 
in the energy range above 1\,GeV to study the spectroscopy of non-strange meson 
resonances \cite{boris} and to search for so-called exotic meson resonances
\cite{exotic}. The very high statistics of the COMPASS experiment allows it to 
continue the event rates with three pions in the final-state even downward to 
the threshold. The (differential) cross sections for the $\pi^-\gamma\to 3\pi$ 
reactions in the low-energy region offer new possibilities to test the strong 
interaction dynamics of the pions as predicted by chiral perturbation theory. 
In a recent work \cite{dreipion} the production amplitudes for $\pi^- \gamma 
\to \pi^-\pi^0 \pi^0$ and $\pi^- \gamma \to \pi^+\pi^-\pi^-$ have been calculated
analytically at one-loop order in chiral perturbation theory. It has been 
found that the next-to-leading order corrections from chiral loops and 
counterterms enhance sizeably (by a factor $1.5 -1.8$) the total cross section 
for neutral pion-pair production $\pi^-\gamma \to\pi^-\pi^0\pi^0$. By contrast
the total cross section for charged pion-pair production $\pi^-\gamma \to\pi^+
\pi^-\pi^-$ remains almost unchanged in comparison to its tree-level result. 
This different behavior can be understood from the varying influence of the
chiral corrections on the pion-pion final-state interaction ($\pi^+\pi^- \to
\pi^0 \pi^0$ versus $\pi^-\pi^- \to  \pi^-\pi^-$).  

The purpose of the present paper is to further improve the theoretical
description of the $\pi^- \gamma \to 3\pi$ reactions by considering the
corresponding QED radiative corrections. We restrict ourselves here to the
simpler case of neutral pion-pair production $\pi^- \gamma \to\pi^-\pi^0\pi^0$, 
for which the number of contributing one-photon loop diagrams is limited to 
about a dozen. Another fortunate circumstance is that the (leading-order) 
chiral $\pi^+\pi^-\to \pi^0\pi^0$ contact-vertex factors out of all photon-loop
diagrams and therefore the radiative corrections to $\pi^- \gamma\to\pi^-\pi^0
\pi^0$ can be represented simply by a multiplicative correction factor $R
\sim \alpha/2\pi$. Infrared finiteness of these virtual radiative corrections 
is achieved (in the standard way) by including soft photon radiation below an 
energy cut-off $\lambda$. Taking  $\lambda=5\,$MeV, we find that the radiative 
corrections to the total cross section for $\pi^- \gamma\to\pi^-\pi^0\pi^0$ vary
between $+2\%$ and $-2\%$ for center-of-mass energies from threshold up to 
$7m_\pi$. An electromagnetic counterterm (necessary in order to cancel all 
ultraviolet divergences generated by the photon-loops) gives an additional 
constant contribution of about $1\%$, however with a large uncertainty. The 
radiative corrections to the charged pion-pair production process $\pi^- \gamma
\to\pi^+ \pi^-\pi^-$ can be roughly estimated to be a factor $2-4$ times larger,
arguing that in this case twice as many charged pions are involved in virtual 
photon-loops and soft photon bremsstrahlung.

\section{Evaluation of one-photon loop diagrams}   
In this section we calculate analytically the radiative corrections to the 
neutral pion-pair photoproduction process: 
\begin{equation}\pi^-(p_1)+\gamma(k,\epsilon\,) \to\pi^-(p_2) +\pi^0(q_1)+ 
\pi^0(q_2)\,,\end{equation}  
as they arise from one-photon loop diagrams at order $\alpha$. For a concise
presentation of our analytical results it is convenient to introduce the 
following dimensionless Mandelstam variables:
\begin{equation} s=(p_1+k)^2/m_\pi^2\,, \quad t=(p_1-p_2)^2/m_\pi^2\,, \quad 
u=(p_2-k)^2/m_\pi^2\,,\end{equation}
with $m_\pi =139.570\,$MeV the charged pion mass. In this (adapted) notation 
$\sqrt{s}\,m_\pi$ is the total center-of-mass energy of the process. We will
also use  frequently the linear combination: 
\begin{equation} \Sigma = s+t+u-2 =(q_1+q_2)^2/m_\pi^2\,,\end{equation}
related to the squared invariant mass of the produced neutral pion-pair.
In the physical region the following inequalities hold: $s>(1+2\sqrt{r_0})^2$, 
$t<0$, $u<0\,$\footnote{The inequality $u<1$ follows immediately from the
definition of $u$. In order to derive the sharper upper bound $u<0$, one uses
the relation for $u$ written in eq.(21) and inserts $y_{\rm max}=1$ and 
$\omega_{\rm max} = (s+1-4r_0)/2\sqrt{s}$. In the end the condition $r_0>1/4$ 
turns out to be crucial for $u$ to take on only negative values.} and 
$4r_0< \Sigma <(\sqrt{s}-1)^2$ where $r_0=(m_{\pi^0}/m_\pi)^2=0.93526$ denotes
the  squared ratio between the neutral pion mass $m_{\pi^0}=134.977\, $MeV and 
the charged pion mass $m_\pi$.

Let us recall the dynamical description of the process $\pi^- \gamma \to \pi^-
\pi^0\pi^0$ at low energies \cite{picross,dreipion}. When choosing for the 
(transversal) real photon $\gamma(k,\epsilon\,)$ the Coulomb-gauge in the 
center-of-mass frame, the conditions $\epsilon \cdot p_1 =\epsilon \cdot k= 0$ 
imply that all diagrams for which the photon couples to the in-coming pion 
$\pi^-(p_1)$ vanish identically. Furthermore, in the convenient parametrization
of the special-unitary matrix-field $U =\sqrt{1-\vec  \pi^{\,2} /f_\pi^2}+ i\vec 
\tau \cdot \vec \pi/f_\pi$ no $\gamma 4\pi$
and $2\gamma 4\pi$ contact-vertices exist (at leading order). Under these 
assumptions one is left with one single $u$-channel pole diagram in which the 
chiral $\pi^+\pi^-\to \pi^0\pi^0$ contact-vertex is followed by a photon-pion 
coupling proportional to $\epsilon\cdot p_2$.

The virtual radiative corrections to $\pi^- \gamma \to \pi^-\pi^0\pi^0$ are
obtained by dressing this tree diagram with a photon-loop in all possible ways 
(see Figs.\,1-4). A fortunate circumstance is that the (leading order) chiral 
$\pi^+\pi^-\to \pi^0\pi^0$ transition amplitude $[(q_1+q_2)^2-m_{\pi^0}^2]/f_\pi^2$ 
depends only on the $\pi^0\pi^0$ invariant mass and thus factors out of all 
photon-loop diagrams. The Coulomb-gauge ($\epsilon\cdot p_1=\epsilon\cdot k=0$)
leaves the scalar product $\epsilon\cdot p_2$ as the only possible coupling 
term for the external real photon. As a consequence of these features the 
radiative corrections due to photon-loops can represented simply by a 
multiplicative correction factor. Its real part which is only of relevance is 
denoted by $R(s,t,u)$. We use dimensional regularization to treat both 
ultraviolet and infrared divergences (where the latter are caused by the 
masslessness of the photon). Divergent pieces of one-loop integrals show up in 
form of the composite constant:
\begin{equation} \xi = {1\over d-4}+{1\over 2}(\gamma_E-\ln 4\pi) +\ln{m_\pi
\over \mu}\,, \end{equation}
containing a simple pole at $d=4$ and $\mu$ is an arbitrary mass scale.
Ultraviolet (UV) and infrared (IR) divergences are distinguished by the
feature of whether the condition for convergence of the $d$-dimensional
integral is $d<4$ or $d>4$. We discriminate them in the notation by putting
appropriate subscripts, i.e. $\xi_{UV}$ and $\xi_{IR}$. In order to simplify
all calculations we employ the Feynman gauge, where the photon propagator is 
directly proportional to the Minkowski metric tensor $g_{\mu\nu}$. We can now
enumerate the analytical expressions for $R(s,t,u)$ as they emerge from the
eight classes of contributing one-photon loop diagrams.

\begin{figure}\begin{center}
\includegraphics[scale=1.,clip]{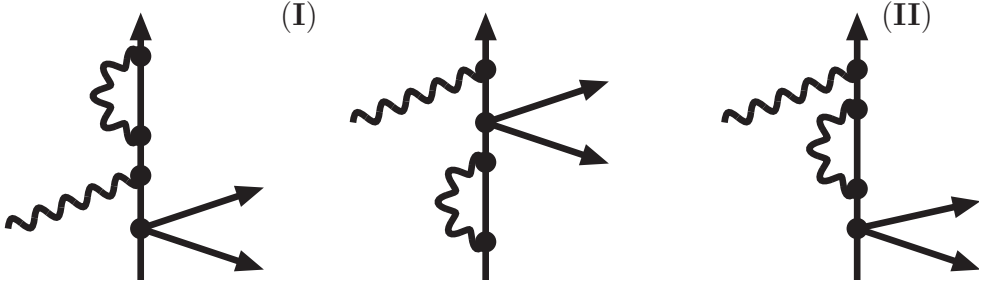}
\end{center}
\vspace{-.5cm}
\caption{One-photon loop diagrams (I) and (II) for neutral pion-pair production 
$\pi^-\gamma \to \pi^-\pi^0\pi^0$. Arrows indicate out-going pions.}
\end{figure}

The two diagrams of class (I) shown in Fig.\,1 introduce the wavefunction
renormalization factor $Z_2-1$ of the pion \cite{comptcor}:   
\begin{equation}R^{(\rm I)}={\alpha \over \pi}\big(\xi_{IR}-\xi_{UV}\big) \,. 
\end{equation}
Diagram (II) involves the once-subtracted (off-shell) selfenergy of the pion
and leads to the result:
\begin{equation}R^{(\rm II)}={\alpha \over \pi}\bigg[-\xi_{UV}+1 -{u+1 \over  2u}
\ln(1-u)\bigg] \,. \end{equation}
\begin{figure}
\begin{center}
\includegraphics[scale=1.,clip]{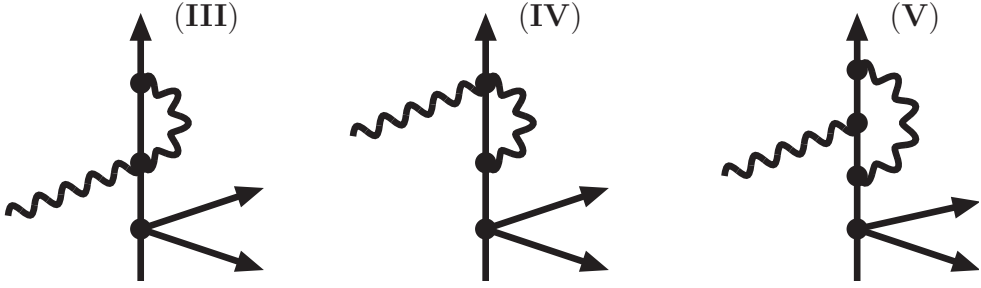}
\end{center}
\vspace{-.5cm}
\caption{One-photon loop diagrams (III), (IV) and (V).}
\end{figure}
Diagram (III) shown in Fig.\,2 gives rise a constant vertex correction:
   \begin{equation}R^{(\rm III)}={\alpha \over 8\pi}\big(6\xi_{UV}-7\big) \,, 
\end{equation}
while diagrams (IV) and (V) generate $u$-dependent vertex corrections:
\begin{equation}R^{(\rm IV)}={\alpha \over 8\pi}\bigg[6\xi_{UV}-6 -{1\over u}+{u-1
\over u^2}(3u+1)\ln(1-u)\bigg] \,, \end{equation}
\begin{equation}R^{(\rm V)}={\alpha \over 8\pi}\bigg[-4\xi_{UV}+5 +{1\over u}+{u^2
+6u+1\over u^2} \ln(1-u)\bigg] \,. \end{equation}
It is astonishing that the last four contributions $R^{(\rm II)}+R^{(\rm  III)}+
R^{(\rm  IV)}+R^{(\rm V)}=0$ sum to zero.
\begin{figure}
\begin{center}
\includegraphics[scale=1.,clip]{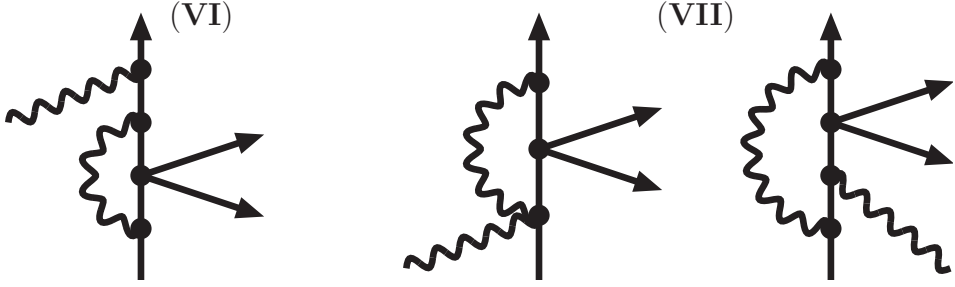}
\end{center}
\vspace{-.5cm}
\caption{One-photon loop diagrams (VI) and (VII).}
\end{figure}

The reducible $u$-channel pole diagram (VI) shown in Fig.\,3 includes a 
photonic vertex correction around the $2\pi^0$ emission vertex. One finds for
its contribution to the (real) $R$-factor the following result: 
\begin{eqnarray} R^{(\rm VI)} &=& {\alpha \over 2\pi}\bigg\{-\xi_{UV}+1+{1-u\over
2u} \ln(1-u) +\sqrt{\Sigma-4\over \Sigma}\ln{\sqrt{\Sigma-4}+\sqrt{\Sigma}\over
2} \nonumber \\ &&+ \bigg(s+t+{u-7\over 2}\bigg) -\!\!\!\!\!\!\int_0^1 dx 
{\ln|x^{-1}+\Sigma(x-1)|-\ln(1-u)\over 1+(u-1)x+\Sigma\, x(x-1)} \bigg\} \,.
\end{eqnarray}
The integrand of the principal-value integral $-\!\!\!\!\!\int_0^1 dx$ has 
simple poles at $x_\pm=[\Sigma+1-u\pm\sqrt{(\Sigma+1-u)^2-4\Sigma}]/2\Sigma$,
but in the physical region $u<0$, $\Sigma >4r_0$ only the pole at $x_-$ lies
inside the unit-interval $0<x<1$. Due to this property an accurate numerical 
treatment of the principal-value integral in eq.(10) (in combination with
further three-body phase space integrations, see eq.(20)) is easily manageable.
We have also checked the Feynman parameter representation of the last loop 
integral in eq.(10) against its  dispersion relation representation: 
\begin{eqnarray}&&  -\!\!\!\!\!\!\int_0^1\!\!dx{\ln|x^{-1}+\Sigma(x-1)|-\ln(1-u) 
\over 1+(u-1)x+\Sigma\, x(x-1)}=  -\!\!\!\!\!\!\int_4^\infty \!\! \!{dx \over x-
\Sigma }\,{2 \over \sqrt{(x+1-u)^2-4x}} \nonumber \\ && \times \ln{\sqrt{x}
(x-u-3)+\sqrt{(x-4)[(x+1-u)^2-4x]} \over 2(1-u)}\,, \end{eqnarray}
where the imaginary part on the right hand side has been calculated via the 
Cutkosky cutting rule. The analytical continuation of the last logarithmic 
term in the  first line of eq.(10) for $0<\Sigma <4$ is $\sqrt{4/\Sigma-1}
\arcsin(\sqrt{\Sigma}/2)$.

In the sum $R^{(\rm I)}+R^{(\rm VI)}$ obtained so far the ultraviolet divergences 
$\xi_{UV}$ do not cancel and the remaining classes of diagrams (VII) and (VIII) 
are actually ultraviolet convergent. In a non-renormalizable effective field 
theory like chiral perturbation theory such a behavior of the radiative
corrections is generic. In order to eliminate all ultraviolet divergences from 
photon-loops additional electromagnetic counterterms have to be introduced 
\cite{knecht}. The black square in the right tree diagram of Fig.\,4 symbolizes
the pertinent electromagnetic counterterm for $\pi^+\pi^- \to \pi^0\pi^0$ 
scattering. It gives rise to the following constant contribution to the 
$R$-factor:      
\begin{equation}R^{(\rm ct)}={3\alpha \over 2\pi}\big(\xi_{UV}+ \bar k\big) \,, 
\end{equation}
where $\bar k$ denotes the finite part of the electromagnetic counterterm
which remains after canceling the ultraviolet divergence $\xi_{UV}$. A 
numerical estimate of $\bar k$ will be given in section 4. For the sake of
completeness we quote the expression for the on-shell $\pi^+(q_+)+\pi^-(q_-)
\to \pi^0(q_1) +\pi^0(q_2)$ scattering amplitude with inclusion of radiative
corrections \cite{knecht}: 
\begin{eqnarray} T_{+-,00} &=&{m_\pi^2 \over f_\pi^2}(\Sigma -r_0)\Bigg\{1+{\alpha  
\over 2\pi} \bigg[ 2(\xi_{IR}-\xi_{UV}) + 3(\xi_{UV}+\bar k) \nonumber \\ && 
-\xi_{UV} +1 +\sqrt{{\Sigma -4\over \Sigma}} \bigg(\ln{\sqrt{\Sigma-4}+\sqrt{
\Sigma}\over 2} -{i \, \pi \over 2}\,\theta(\Sigma -4) \bigg) \nonumber \\ && 
+(\Sigma -2) \int_4^\infty\!\!{dx\over x-\Sigma -i0^+} \, {2\xi_{IR}+\ln(x-4) \over
\sqrt{x^2-4x}} \bigg] \Bigg\} \,,  \end{eqnarray}
where $\Sigma  = (q_1+q_2)^2/m_\pi^2$ and $f_\pi = 92.4\,$MeV denotes the pion
decay constant. In the order given the terms in the square bracket correspond 
to the pion wavefunction renormalization factor $Z_2-1$, the electromagnetic 
counterterm, and the one-photon exchange contribution. Note that we have used 
the (concise) spectral function representation for the infrared divergent 
(scalar) loop integral involving one photon and two pion propagators. If an
infinitesimal photon mass $m_\gamma$ is introduced as an (alternative) infrared 
regulator the infrared divergence $\xi_{IR}$ is to be identified with the
logarithm $\ln(m_\pi/m_\gamma)$.

Next, we come to the irreducible $s$-channel diagrams of class (VII) shown in
Fig.\,3. Since the contribution from the left diagram (involving the two-photon
contact-vertex) gets (partly) canceled by a term from the right diagram it is 
advantageous to specify only their total contribution to the $R$-factor. After
reducing the loop integrals with four propagators one obtains the following
result: 
\begin{eqnarray} R^{(\rm VII)} &=& {\alpha \over 2\pi}(u-1)\Bigg\{{D(t)-D(\Sigma)
\over 2(\Sigma-t)} +\int_0^1 dx {1-2x+2x^2 \over 1-t\, x(1-x)}  \nonumber \\ &&
\times {\ln|x^{-1}+\Sigma(x-1)|-\ln(s-1)\over 1+(s-1)x+\Sigma\,x(x-1)} \Bigg\} 
\,.\end{eqnarray}
Note that the last denominator has no poles in the physical region, since
$s-1+\Sigma(x-1) >s-1-\Sigma>s-1-(\sqrt{s}-1)^2=2(\sqrt{s}-1)>0$. By taking
the absolute magnitude of the arguments of logarithms one gets directly a  
suitable representation of the only relevant real part. It is a fortunate
circumstance that the Feynman-parameter representation of loop functions leads
to expressions which can be handled easily numerically in the physical region.

\begin{figure}
\begin{center}
\includegraphics[scale=1.,clip]{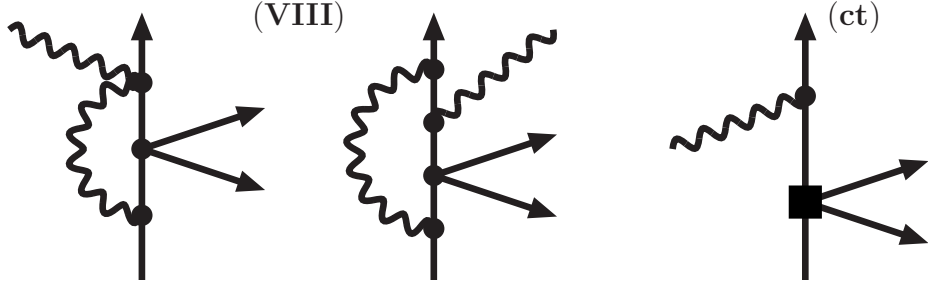}
\end{center}
\vspace{-.5cm}
\caption{One-photon loop diagrams (VIII). The black square in the right tree 
diagram (ct) symbolizes the electromagnetic counterterm for $\pi^+\pi^-\to 
\pi^0\pi^0$ scattering.}
\end{figure}

Finally, we come to the irreducible $u$-channel diagrams of class (VIII) shown 
in Fig.\,4. The contribution from the left diagram gets completely absorbed by
a term from the right diagram. The resulting contribution to the (real) 
$R$-factor includes an infrared divergent term with a non-trivial 
$t$-dependence and after putting all pieces together it reads:   
\begin{eqnarray} R^{(\rm VIII)} &=& {\alpha \over 2\pi}\Bigg\{{1-u\over  2}\bigg[
{D(t)-D(\Sigma)\over \Sigma-t}- {1 \over u}\ln(1-u)  \nonumber \\ && + 
-\!\!\!\!\!\!\int_0^1 dx {1+4x^2-t\,x(1+x)\over 1-t\,x(1-x)}\,{\ln|x^{-1}+\Sigma
(x-1)|-\ln(1-u)\over 1+(u-1)x+\Sigma\,x(x-1)} \bigg]\nonumber \\ &&+{t-2 \over 
\sqrt{t^2-4t}} \bigg[4\Big(\xi_{IR}+\ln(1-u)\Big) \ln{\sqrt{4-t}+\sqrt{-t}\over
2} +{\rm Li}_2(w)\nonumber \\ && -{\rm Li}_2(1-w)+{1\over 2}\ln^2 w-{1\over 2}
\ln^2(1-w) + {\rm Li}_2(h_-)-{\rm Li}_2(h_+)\bigg]\nonumber \\ &&  +(2-t)
\int_0^1 dx{\ln|1+ \Sigma\, x(x-1)|\over 1-t\,x(1-x)} \Bigg\} \,,\end{eqnarray}
with the abbreviations
\begin{equation} w = {1\over 2}\Bigg(1-\sqrt{-t\over 4-t}\,\Bigg)\,, \qquad
  h_\pm = {1\over 2}\Big(t \pm \sqrt{t^2-4t}\, \Big)\,. \end{equation}
One observes that the term proportional to $D(t)-D(\Sigma)$ drops out in the 
sum  $R^{(\rm VII)} + R^{(\rm VIII)}$ and therefore we do not need to specify it. 
Li$_2(w) = \sum_{n=1}^\infty n^{-2} w^n= w\int_1^\infty dx [x(x-w)]^{-1} \ln x$ denotes 
the conventional dilogarithmic function. Several of the results derived in 
section 3 of ref.\cite{bremscor} have been useful in order to obtain the 
expression for $R^{(\rm VIII)}$  written in eq.(15).
  
\section{Infrared finiteness}
In the next step we have to consider the infrared divergent terms proportional
to $\xi_{IR}$ present in eqs.(5,15). At the level of the measurable cross 
section these get eliminated by contributions from (undetected) soft photon 
bremsstrahlung. In its final effect, the (single) soft photon radiation off 
the in- or out-going $\pi^-$ multiplies the tree-level differential cross 
section for $\pi^-\gamma \to\pi^- \pi^0\pi^0$ by a (universal) factor 
\cite{comptcor,bremscor}:  
\begin{equation} \delta_{\rm soft}=2R_{\rm soft}= \alpha\, \mu^{4-d}\!\!\int\limits_{
|\vec l\,|<\lambda} \!\!{d^{d-1}l  \over (2\pi)^{d-2}\, l_0} \bigg\{ {2p_1\cdot p_2 
\over p_1 \cdot l \, p_2 \cdot l} - {m_\pi^2 \over (p_1 \cdot l)^2} - {m_\pi^2 
\over (p_2 \cdot l)^2} \bigg\} \,, \end{equation}
which depends on a small energy cut-off $\lambda$. Working out this momentum 
space integral by the method of dimensional regularization (with $d>4$) one 
finds the following contribution from soft photon emission to the $R$-factor: 
\begin{eqnarray}R_{\rm soft}^{(\rm cm)}&=& {\alpha \over 4\pi}\Bigg\{4\bigg[1+{2 t-4 
\over\sqrt{t^2-4t}} \ln{\sqrt{4-t}+\sqrt{-t}\over 2}\bigg] \bigg(\ln{m_\pi\over
2\lambda} -\xi_{IR}\bigg) \nonumber \\ && + {s+1 \over s-1} \ln s + {2\omega
\over \sqrt{\omega^2-1}}\ln\Big(\omega+\sqrt{\omega^2-1}\,\Big)+(t-2) \nonumber 
\\ && \times \int_0^1 dx {s+1-\Sigma\,x\over [1-t\,x(1-x)] \sqrt{W}} \ln{ s+1
-\Sigma\,x+\sqrt{W} \over s+1-\Sigma\,x-\sqrt{W}}\Bigg\} \,,\end{eqnarray}
with the abbreviation $W= (s+1-\Sigma\,x)^2-4s[1-t\,x(1-x)]$. In order to
simplify the last term in eq.(18) we have made use of the relation $\Sigma= 
s+1-2 \omega\sqrt{s}$, where $\omega$ denotes the center-of-mass energy of the
out-going negative pion $\pi^-(p_2)$ divided by $m_\pi$. Note that the terms 
beyond those proportional to $\ln(m_\pi/2\lambda) -\xi_{IR}$ are specific for 
the evaluation of the soft photon correction factor $R_{\rm soft}$ in the 
center-of-mass frame with $\lambda$ an infrared cut-off therein.  

In order to present a concrete example we have evaluated the complete radiative
correction factor $R$ at the threshold in the isospin limit: $s_{\rm th}=9$, 
$t_{\rm  th}=-4/3$,  $u_{\rm th}=-5/3$, $\Sigma_{\rm th}=4$,  $\omega_{\rm  th}=1$. In 
this case one gets numerically:
\begin{equation}R_{\rm th}={\alpha \over 2\pi} \bigg\{11.093+3\bar k+ \bigg(2-
{5\over  2}\ln 3\bigg)\ln{m_\pi \over 2\lambda} -0.725\bigg\} \,, \end{equation} 
where the terms in the curly bracket correspond in the order written to virtual
photon-loops, the electromagnetic counterterm, the universal soft photon 
contribution, and the  soft photon contribution specific for imposing an 
infrared cut-off via $|\vec l\,|<\lambda$ in the center-of-mass frame.  
\section{Results: radiative corrections to cross sections}
After inclusion of radiative corrections the total cross section for neutral 
pion-pair production $\pi^-\gamma \to\pi^-\pi^0\pi^0$ depends also on
the infrared cut-off $\lambda$ for undetected soft photons. We multiply the 
squared tree-level amplitude by $1+2R(s,t,u,\lambda)$ and integrate over the 
three-pion phase space. Applying the usual flux and symmetry factors the
total cross  section reads:
\begin{eqnarray} \sigma_{\rm tot}(s,\lambda) &=& {\alpha\,m_\pi^2\over 32\pi^2f_\pi^4
(s-1)}\int_{1}^{\omega_{\rm  max}}\!\!d\omega\,(\omega^2-1)^{3/2}\sqrt{\Sigma-4r_0 \over 
\Sigma }\nonumber \\ && \times\int_{-1}^1 dy\,(1-y^2) \bigg({\Sigma-r_0\over
u-1} \bigg)^{\!\!2}\,\Big[1+2R(s,t,u,\lambda)\Big] \,,
\end{eqnarray}
with $\omega_{\rm max} = (s+1-4r_0)/2\sqrt{s}$ the endpoint energy of the 
out-going $\pi^-$ divided by $m_\pi$. Using the relations $\Sigma= s+t+u-2= 
s+1-2 \omega \sqrt{s}$ and:
\begin{equation} u= 1+{1-s \over\sqrt{s}}\Big(\omega-y  \sqrt{\omega^2-1}\,\Big)
\,, \qquad t= 2-(s+1){\omega \over\sqrt{s}}+{1-s \over\sqrt{s}} y \sqrt{
\omega^2-1}\,, \end{equation} 
valid in the center-of-mass frame the whole integrand in eq.(20) becomes a
function of $\omega$ and the directional cosine $y$.

\begin{figure}
\begin{center}
\includegraphics[scale=.5,clip]{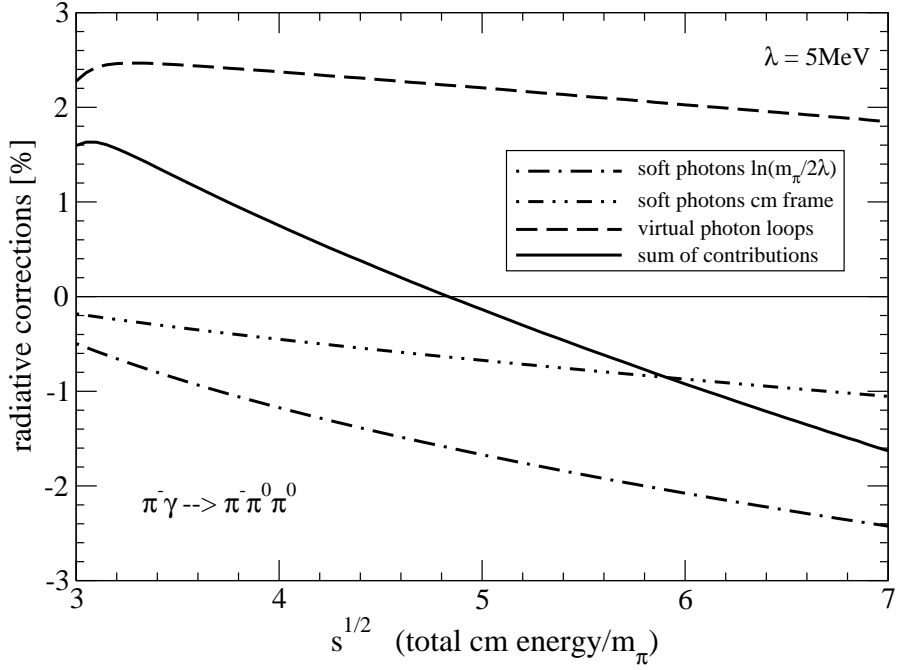}
\end{center}
\vspace{-0.6cm}
\caption{Radiative corrections to the total cross section for neutral 
pion-pair photoproduction $\pi^- \gamma \to \pi^-\pi^0\pi^0$ as a function of 
the center-of-mass energy $\sqrt{s}\,m_\pi$. The infrared cut-off for soft 
photon emission has been set to the value $\lambda = 5\,$MeV.} 
\end{figure}

Fig.\,5 shows in percent the radiative corrections to the total cross section 
for neutral pion-pair production $\pi^-\gamma \to\pi^-\pi^0\pi^0$ as a function 
of the center-of-mass energy $\sqrt{s}\,m_\pi$. The dashed-dotted and dashed 
curves display the separate contributions from soft photon bremsstrahlung and 
virtual photon-loops. In each case the radiative correction is calculated as 
the ratio of the shift in $\sigma_{\rm tot}(s,\lambda)$ induces by the respective 
$R$-factor divided by the tree-level cross section. As in ref.\cite{comptcor} 
the infrared cut-off $\lambda$ for undetected soft photons has been set to 
$\lambda=5\,$MeV, a value which seems appropriate for the COMPASS experiment. 
The full line in Fig.\,5 shows the complete radiative corrections. One observes 
an almost linear decrease which ranges from $+1.6\%$ at threshold to $-1.6\%$ 
at the center-of-mass energy $\sqrt{s}\,m_\pi=7m_\pi$. An interesting feature
is that the positive radiative corrections from the virtual photon-loops get 
gradually reduced and turned into negative values by the soft photon
contributions.   
    
The finite part of the electromagnetic counterterm $\bar k$ shifts the 
radiative corrections (displayed by the full curve in Fig.\,5) by a 
constant amount of $3\alpha\bar k/\pi= 0.7\%\cdot \bar k$. In order to give an
estimate for $\bar k$ we exploit the elaborate result of ref.\cite{pionium}
for the pionium decay amplitude $(a_0-a_2)m_\pi+\varepsilon$. Guided by eq.(13) 
we identify $(\alpha/ 2\pi)(3\bar k +1)$ with the ratio $\varepsilon^{\rm  elm}/(
a_0-a_2)$, where $\varepsilon^{\rm  elm}$ is the electromagnetic correction to 
the pionium decay amplitude. Subtracting from $\varepsilon= (6.1\pm 1.6)\cdot 
10^{-3}$ the contribution $4.8\cdot 10^{-3}$ due to the (charged and neutral)
pion mass difference (see eqs.(4.28,4.29) in ref.\cite{pionium}) one gets 
$\varepsilon^{\rm elm}= (1.3\pm 1.6)\cdot 10^{-3}$. Together with the leading order 
expression for the $\pi\pi$-scattering length difference $a_0-a_2 = 9m_\pi/
(32\pi f_\pi^2) = 0.204 m_\pi^{-1}$ one arrives at the estimate $\bar k = 1.5\pm 
2.2$ for the electromagnetic counterterm. Its central value implies a constant 
shift of the radiative corrections to $\pi^-\gamma \to\pi^-\pi^0\pi^0$ by
about $1.0\%$. The large errorbar of $\bar k = 1.5\pm 2.2$ introduces at the
same time a wide errorband to the full curve in Fig.\,5. Still an allowed
option is to neglect to electromagnetic counterterm, setting $\bar k=0$.

\begin{figure}
\begin{center}
\includegraphics[scale=.5,clip]{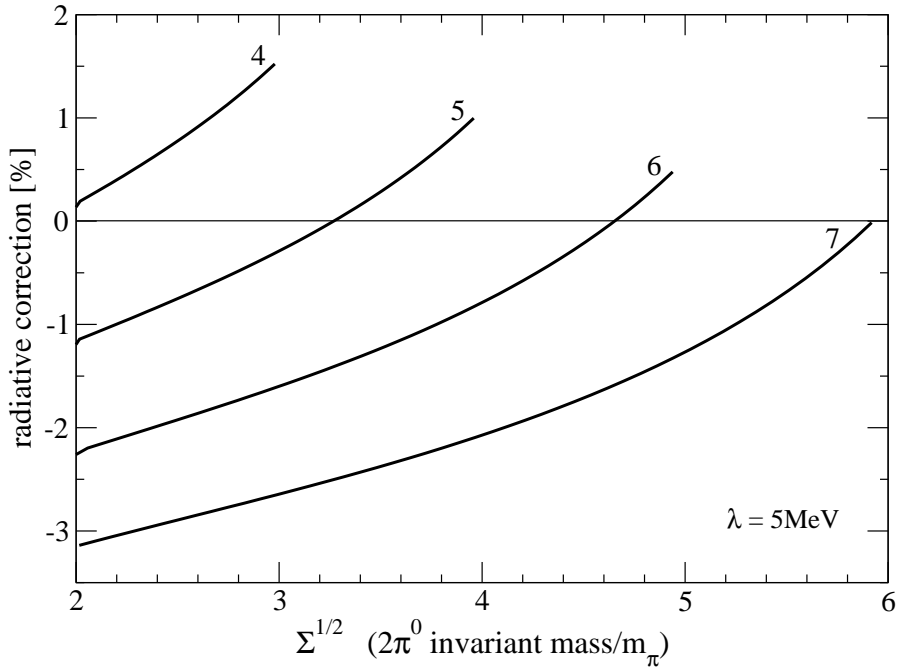}
\end{center}
\vspace{-0.6cm}
\caption{Radiative corrections to the $\pi^0\pi^0$ mass spectra for neutral 
pion-pair production $\pi^- \gamma \to \pi^-\pi^0\pi^0$ as a function of 
the $\pi^0\pi^0$ invariant mass $\sqrt{\Sigma}\,m_\pi$. The numbers on the
curves correspond to $\sqrt{s}$.} 
\end{figure}

Finally, we consider radiative corrections to more exclusive observables. An
obvious candidate is the $\pi^0\pi^0$ mass spectrum $d\sigma/dm_{00}$ with $m_{00} 
=\sqrt{\Sigma}\,m_\pi$ the $\pi^0\pi^0$ invariant mass. The differential cross 
section $d\sigma/dm_{00}$ is obtained by omitting the $d\omega$-integration in 
eq.(20) and applying the normalization factor $ m_\pi^{-1}\sqrt{\Sigma/s}$. 
Fig.\,6 shows in percent the radiative corrections to the $\pi^0\pi^0$ mass 
spectrum for neutral pion-pair production $\pi^- \gamma \to \pi^-\pi^0\pi^0$ as 
a function of the $\pi^0\pi^0$ invariant mass $\sqrt{\Sigma}\,m_\pi$. The 
numbers (4,\,5,\,6,\,7) on the four rising curves correspond to $\sqrt{s}$, the 
total center-of-mass energy divided by $m_\pi$. The electromagnetic counterterm
$\bar k$ shifts again the whole pattern by a constant $3\alpha \bar k/\pi$. 
 
In summary, we find that the radiative corrections to neutral pion-pair
production $\pi^- \gamma \to \pi^-\pi^0\pi^0$ are comparable in size to those
for pion Compton scattering $\pi^- \gamma \to\pi^- \gamma $ \cite{comptcor}.

\end{document}